\documentclass[final]{aipproc}
\usepackage{graphicx}

\layoutstyle{6x9}

\newcommand\Ms{M$_\odot$}

\begin{document}

\title{Local and Global Radiative Feedback from Population III Star Formation}

\classification{97.20.Wt, 98.35.Ac, 98.62.Ai, 98.70.Vc}
\keywords      {Population III stars, reionization, radiative feedback, 
  high redshift structure formation}

\author{Brian W. O'Shea}{
  address={Department of Physics and Astronomy and Lyman Briggs College, 
Michigan State University,
    East Lansing, MI, USA; oshea@msu.edu}
}

\author{Daniel J. Whalen}{
  address={McWilliams Center for Cosmology, Carnegie Mellon University, 
    Pittsburgh, PA, USA; dwhalen2@cmu.edu}
}

\begin{abstract}
We present an overview of recent work that focuses on understanding the 
radiative feedback
processes that are potentially important during Population III star formation.  
Specifically, we examine the effect of the Lyman-Werner
(photodissociating) background on the early stages of primordial star 
formation, which serves to delay the onset of star formation in a given
halo but never suppresses it entirely.  We also examine the effect that 
both photodissociating and ionizing
radiation in I-fronts from nearby stellar systems have on the formation of
primordial protostellar clouds.  Depending on the strength of the incoming radiation
field and the central density of the halos, Pop III star formation can be suppressed,
unaffected, or even enhanced.  Understanding these and other effects 
is crucial to modeling Population III star formation and to 
building the earliest generations
of galaxies in the Universe.
\end{abstract}

\maketitle

%%%%%%%%%%%%%%%%%%%%%%%%%%%%%%%%%%%%%%%%%%%%
%% MAINMATTER
%%%%%%%%%%%%%%%%%%%%%%%%%%%%%%%%%%%%%%%%%%%%

\section{Introduction}
Population III star formation is generally considered to be a simple problem 
compared to galactic star formation.  This is only strictly true, however, for the
so-called ``Population III.1'' stars, which form 
in the absence of any non-cosmological influences \cite{osheaFS3,McKee08}.  
Our current understanding is that the transition from Pop III to metal-enriched
star formation occurs very quickly locally, but globally this transition extends
over a wide range of redshifts
(possibly down to $z \sim 6$).  As a result, the vast majority of Population 
III stars will form in a universe where earlier generations of primordial and 
metal-enriched stars have produced both photodissociating and ionizing radiation, as 
well as X-ray, cosmic rays, and kinetic feedback.  

This feedback could very possibly 
affect the final mass of the primordial stars, and stars that form under these 
conditions are often referred to as
``Population III.2'' stars. \cite{osheaFS3}  This term is somewhat misleading,
because it implies homogeneity in the effects of these various types of feedback.  This
is incorrect.  Lyman-Werner radiation (E$_\nu \sim 11.18-13.6$~eV)
photodissociates molecular hydrogen, which typically delays star formation.
Ionizing radiation destroys H$_2$ and ionizes hydrogen, which in principle completely
halts star formation; however, massive stars are short-lived, and the huge
free electron populations in relic HII regions can catalyze tremendous
amounts of H$_2$ and HD formation, possibly spurring primordial star formation
and changing the characteristic mass scales of these stars.  X-rays and cosmic rays
most likely also promote the creation of a significant free electron population, catalyzing
H$_2$ and HD formation and accelerating primordial star formation.  To further complicate
matters, a single star (or stellar population) produces multiple types 
of feedback whose net effect
depends heavily on the relative positioning of halos and cosmic epoch.  
One can safely say that the
majority of Population III stars form in situations where the feedback
acting upon them
is context-sensitive and the final outcome is 
ambiguous.  In this paper, we present studies of
the effects of a global photodissociating background on a primordial star-forming
halo, and, separately, multifrequency radiation hydrodynamics simulations of 
impinging ionization fronts on idealized cosmological halos.

\section{Effects of a Global Photodissociating Background}

In order to study the effects of a global photodissociating background, we 
take a single ``standard'' Pop III star formation simulation using the Enzo code \cite{on07, enzo05} 
and gradually increase the strength of the Lyman-Werner
(H$_2$ photodissociating) background.
Other than the strength of the photodissociating background, all other simulation
parameters (as well as the initial conditions) are kept constant.

\begin{figure}
  \includegraphics[height=.5\textheight]{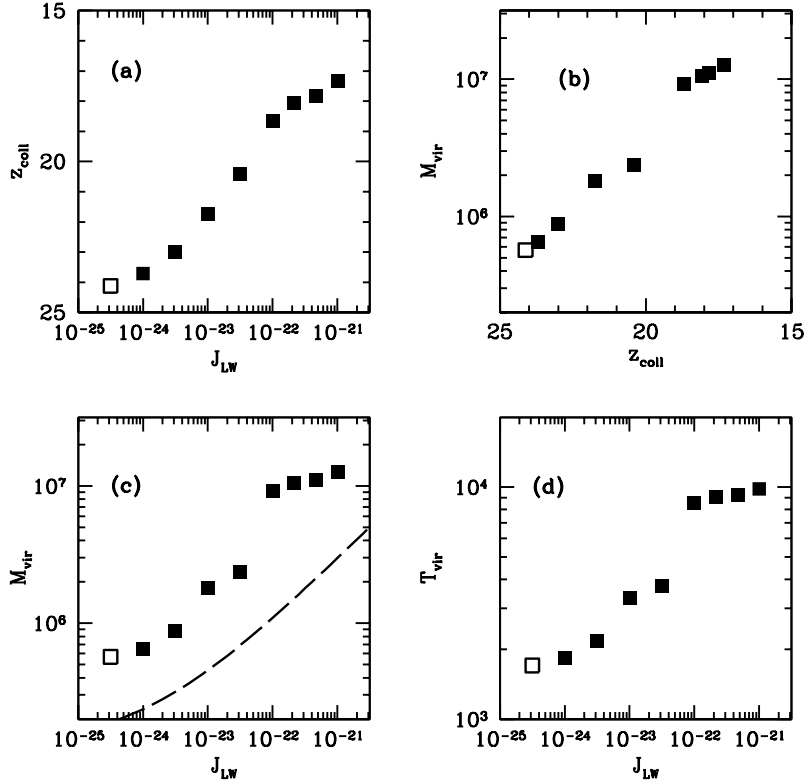}
  \caption{Mean halo quantities for several simulations with the same cosmic 
realization but a range of Lyman-Werner molecular hydrogen
photodissociating flux
backgrounds.  
Panel (a):  J$_{LW}$ vs. halo collapse redshift.
Panel (b): halo virial mass vs. halo collapse redshift.
Panel (c): halo virial mass vs. J$_{LW}$.
Panel (d): halo virial temperature vs. J$_{LW}$.
The J$_{21} = 0$ ``control'' results are shown as an open square 
(and is at log J$_{LW} = -24.5$ in the panels which
are a function of J$_{LW}$).
In the bottom left panel, the dashed line corresponds to 
the fitting function for threshold mass
from \cite{machacek01}, Eqn. 8.
Figure is from \cite{on08}.}
  \label{fig.lweffects}
\end{figure}

\begin{figure}
  \includegraphics[height=.5\textheight]{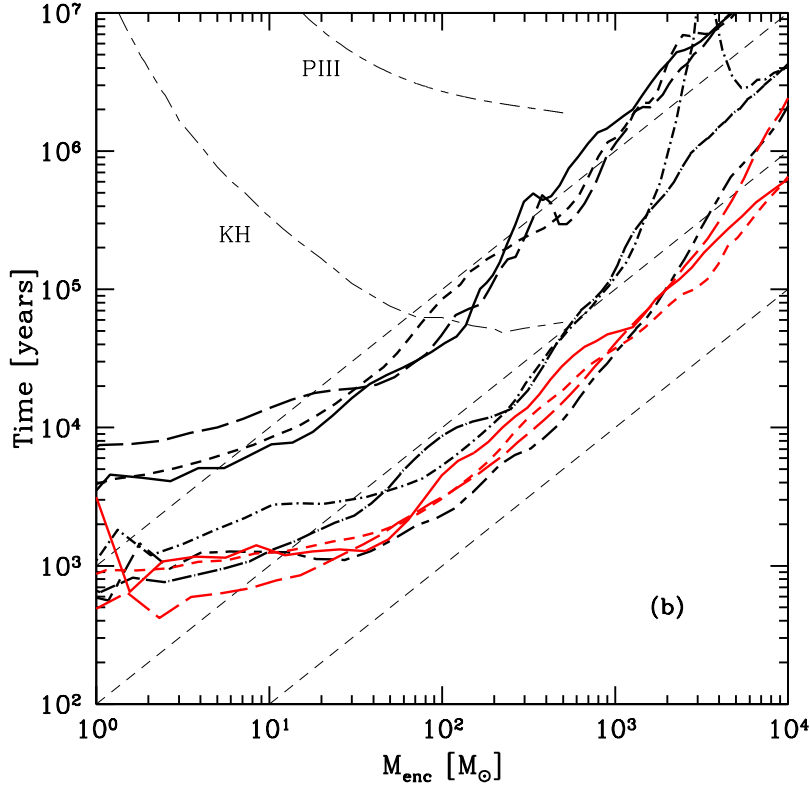}
  \caption{Spherically-averaged, mass-weighted accretion 
time as a function of 
enclosed baryon mass, shown at the point where the maximum
baryon number density in the simulation is approximately $10^{10}$~cm$^{-3}$.
Line types and weights correspond to those in Figure 9  
of \cite{on08} and, from top to bottom, generally go from low to high FUV flux.
The upper and lower light short-long-dashed curves which extend from the
upper left corner correspond to 
the main sequence lifetime of a massive Population III star of that mass and
the Kelvin-Helmholtz time scale of a Population III time scale with a given 
luminosity and
radius.  All values are taken from~\cite{schaerer02}.
The three light diagonal short-dashed lines which extend
from bottom left to top right correspond to masses
accreted using constant accretion rates of (from top to bottom)
$\dot{m} = 10^{-3}$, $10^{-2}$, and $10^{-1}$ M$_\odot/$yr.
Figure is from \cite{on08}.
}
  \label{fig.lwaccretion}
\end{figure}

Figures~\ref{fig.lweffects} and~\ref{fig.lwaccretion} show the primary
results from this work, which are examined in much more detail in~\cite{on08}.
Figure~\ref{fig.lweffects} shows the effect that turning up the FUV background
has on the halo collapse redshift, virial mass of the halo, and virial temperature
of the halo.  In general, raising the strength of the photodissociating background
delays Population III star formation to lower redshifts, when the halo is more
massive and hotter.  The FUV background only delays star formation -- it is never
completely suppressed, because a small amount of molecular hydrogen can always form.
Figure~\ref{fig.lwaccretion} shows the estimated accretion of 
gas onto the protostellar core based on conditions at the time 
where the gas finally collapses to high
densities in each of these simulations.  In general, simulations with larger
FUV backgrounds are hotter, and since the inflow of gas occurs at approximately 
the sound speed in the halo, the accretion rates are higher.  A naive reading
of this figure implies that
Pop III stars in the presence of a FUV background are more massive, but
detailed semi-analytical models suggest that
this may not actually be so  \cite{Tan04,McKee08}. 

\section{Local Effects Of Ionizing Radiation}

\begin{figure}
  \includegraphics[height=.4\textheight]{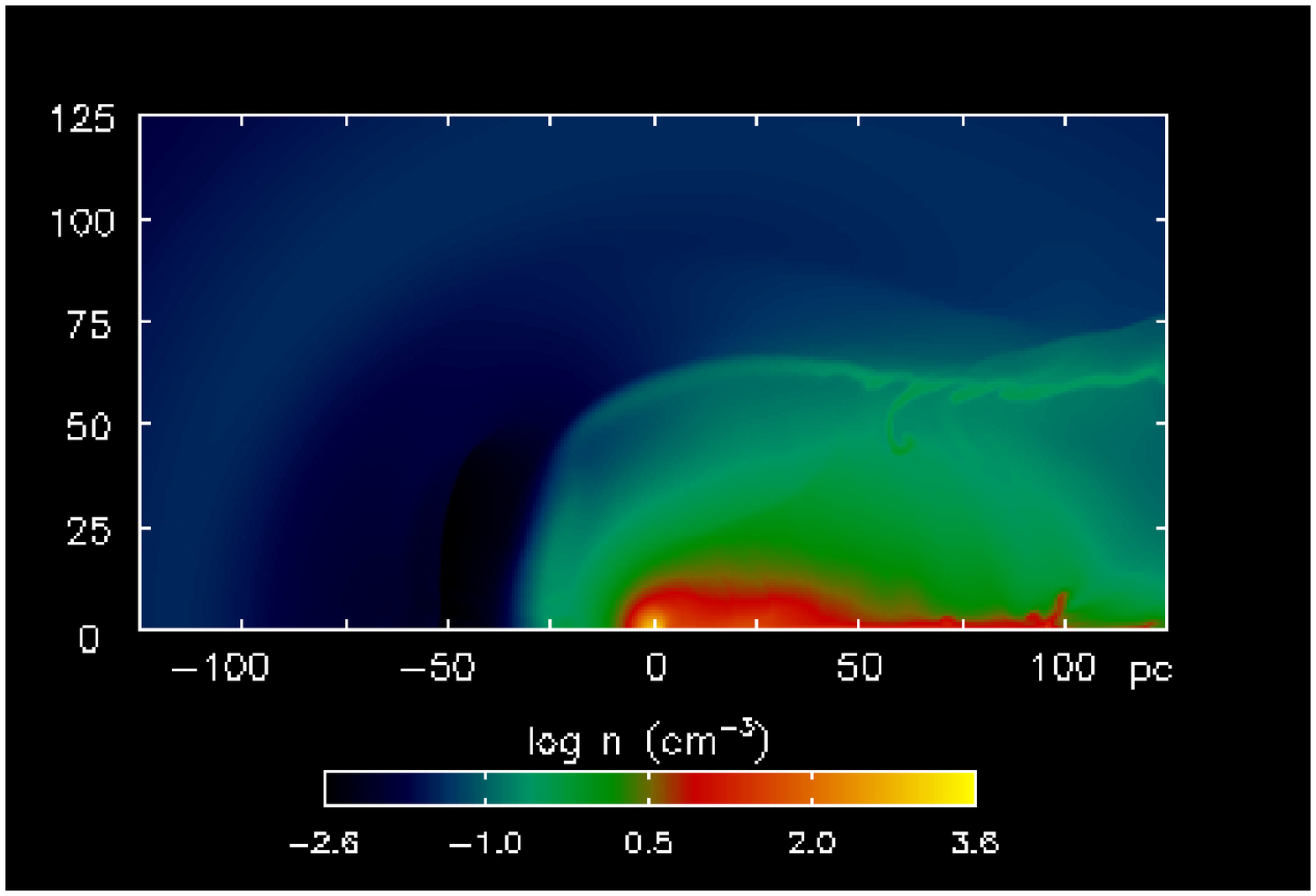}
  \caption{Evaporated halo with $n_c =$ 1596 cm$^{-3}$,  150 
pc from a 60 \Ms \ star at 10 Myr.  The core of the halo is slightly 
displaced to the left of center by backflow from the collapsed shadow 
on the right.  Figure from \cite{whalen10}.}
  \label{fig.hii_image}
\end{figure}

Cosmological simulations \cite{oshea05,yoshida05} suggest
that HII regions from nearby massive stars may delay 
primordial star
formation in a neighboring halo, but the residual electron population after
the death of the original star will ultimately make more H$_2$ and HD 
than would be possible in an undisturbed halo.  The major issue with these
cosmological simulations is that they very loosely approximate I-front 
interactions with gas in neighboring
halos, using static density fields and assuming monochromatic incident radiation.
These two assumptions seriously limit the validity of the results from such
calculations.

In the study presented here (originally published in \cite{whalen08, whalen10}), 
we use idealized simulations that combine 
multifrequency radiation transport with non-equilibrium
chemistry, hydrodynamics, and radiative cooling.  Ionization fronts
have finite widths, with H$_2$-photodissociating radiation and X-rays generally traveling
ahead of the ionizing radiation.  This is particularly true in Population 
III stars, where the surface temperatures are extremely high.  
In this proceedings we present results
from \cite{whalen10}, which focuses on I-fronts from Pop III stars with masses of
15-40 \Ms.  A grid of simulations are performed, varying both the central number density of 
an idealized $1.35 \times 10^5$~\Ms~halo and the distance between the star and the density
peak of this halo (and, thus, the intensity of incident radiation at the halo).

\begin{figure}
  \includegraphics[height=.3\textheight]{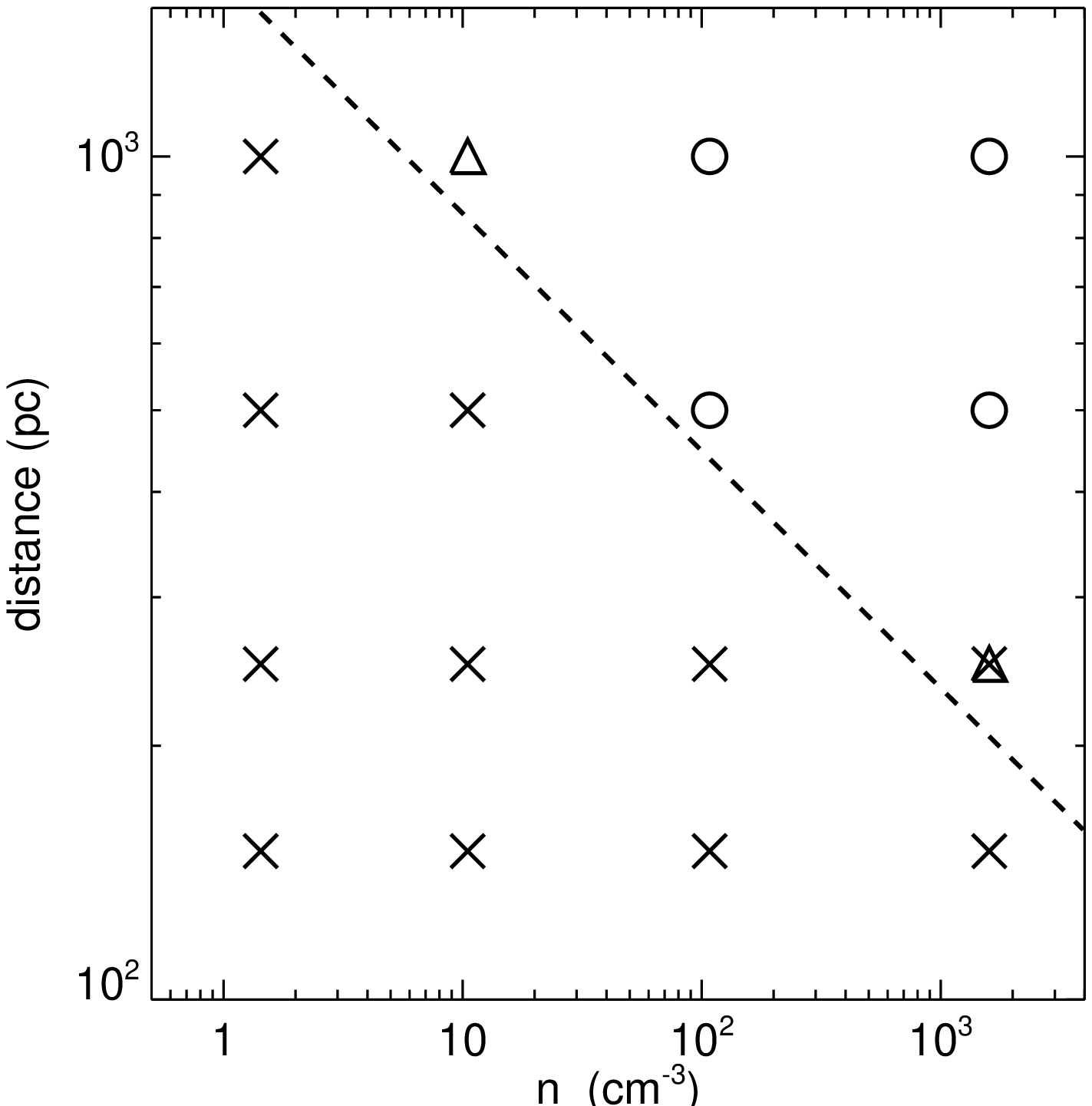}
  \includegraphics[height=.3\textheight]{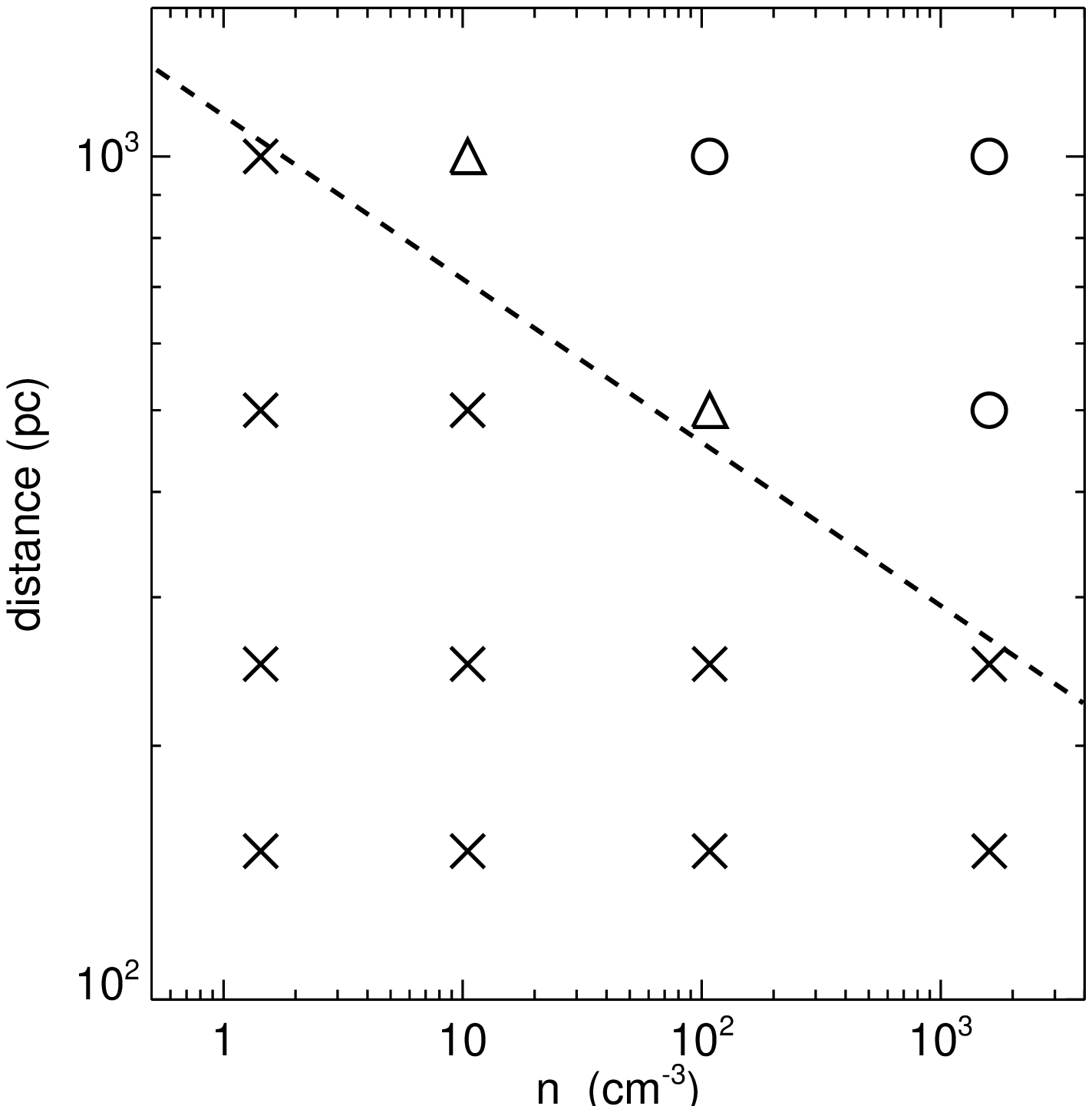}
  \caption{Radiative and kinetic feedback on star formation near a 25 \Ms \ 
star (left) and a 40 \Ms \ star (right). Completely evaporated halos with 
no star formation are labeled by crosses, while halos with delayed or
unchanged star formation are marked by triangles and circles, respectively.  
The triangle overlaid on the cross in the 25 \Ms \ panel signifies that the 
halo can form a delayed star if the SN goes off in the 6.9 $\times$ 10$^5$ 
\Ms \ halo but not if it is in the 2.1 $\times$ 10$^{6}$ \Ms \ halo.  The 
dotted lines again define the boundary for star formation in the evaporated 
halos, above which it proceeds and below which it is quenched.
Figure from \cite{whalen10}.}
  \label{fig.hii_endresult}
\end{figure}

Figure~\ref{fig.hii_image} shows the distribution of baryons at the end of a 
representative simulation with a high central halo density.  In this
case, the I-front has washed over the halo, but H$_2$ has formed in the relatively dense
gas around the halo core and shielded it from the incident Lyman-Werner photons.
Figure~\ref{fig.hii_endresult} shows the outcomes for a suite of models, varying
source distance and halo central density, for two different stellar masses (and thus
spectral shapes).  Halos close to the star and with relatively low central densities are 
evacuated of gas, suppressing star formation.  Halos that are further way or have higher 
central gas densities see either a delay in gas collapse or no effect whatsoever.

The outcomes are much less clear-cut than in the situation where
we examine the effect of a photodissociating background alone -- depending on the central
baryon density and strength of the incoming radiation background, halos
can be completely evaporated, have delayed star formation, or be 
completely unaffected.

\section{Discussion and Conclusions}

The vast majority of primordial stars do not form in splendid
isolation -- the evolution of the protostellar gas clouds that 
will become Population III stars can be strongly influenced by 
radiation created by the 
previous generations of star formation.  Depending on the circumstances,
this feedback can have a variety of outcomes.  It appears that photodissociating
radiation can never completely halt the formation of molecular hydrogen 
in T$_{vir} < 10^4$~K halos, and
can only delay the formation of a primordial star in a given halo, until the halo
has grown sufficiently for the small amount of H$_2$ in it to be an 
efficient coolant.  
FUV background only delays, and never completely suppresses, H$_2$
formation and thus Pop III star formation.
This appears to be a robust result: Machacek et al. \cite{machacek01} suggested
that the FUV background imposes a threshold mass for star formation that rises
 with the strength of the FUV background, and, separately, Wise \& Abel 
\cite{wise07}
show that, even with very high FUV backgrounds, collisional ionization drives H$_2$
formation faster than it can be dissociated.

The interaction of ionization fronts with a primordial halo
can have a variety of effects, from completely evaporating the halo to having no 
influence on its star formation whatsoever.  It is clear that in order to accurately 
investigate these effects one needs multifrequency radiation hydrodynamics simulations.
Furthermore, it seems critical to move beyond the tiny simulation volumes and single
stars that are typically used in Pop III star formation, and to instead simulate
more realistic circumstances:  high-sigma peaks at high redshift (corresponding to clusters of 
halos), and lower-sigma peaks at all redshifts, both with appropriate radiation
backgrounds.  Only when we study a much wider variety of potentially star-forming primordial
halos can we understand Population III star formation and how early
stellar populations regulate their own growth.

%%%%%%%%%%%%%%%%%%%%%%%%%%%%%%%%%%%%%%%%%%%%%%%%
%% BACKMATTER
%%%%%%%%%%%%%%%%%%%%%%%%%%%%%%%%%%%%%%%%%%%%%%%%

\begin{theacknowledgments}
This work was carried out in part under the 
auspices of the National Nuclear Security Administration of the U.S. Department of Energy 
at Los Alamos National Laboratory under Contract No. DE-AC52-06NA25396.  BWO was 
supported by a LANL Director's Postdoctoral Fellowship (DOE LDRD grant 20051325PRD4),
and was later supported by the LANL Institute for Geophysics and Planetary Physics. 
DJW was supported by the McWilliams Fellowship at the Bruce and Astrid McWilliams Center 
for Cosmology at CMU.  Simulations described in this paper were performed at SDSC and NCSA 
with computing time provided by 
NRAC allocation MCA98N020, and at LANL using time provided by the 
Institutional Computing Program.  
%We would like to thank Robert Hueckstaedt, 
%Thomas McConkie, and Michael Norman for allowing us to use figures from papers
%where they are co-authors.
\end{theacknowledgments}



\bibliographystyle{aipproc}   % if natbib is available

\end{document}